\documentclass[aps,twocolumn,nofootinbib]{revtex4-2}
\usepackage{mathtools}
\usepackage{booktabs}
\usepackage{adjustbox} 
\usepackage{amssymb}
\usepackage{array}
\usepackage{makecell}  
\usepackage{graphicx}
\usepackage{dcolumn}
\usepackage{bm}
\usepackage{graphicx}    
\usepackage{subcaption}  
\usepackage{enumitem}   
\usepackage{titlesec}
\titlespacing*{\subsection}
  {0pt}   
  {0.8em} 
  {0.3em} 

\usepackage{hyperref}
\hypersetup{
    colorlinks,
    citecolor=blue,
    filecolor=blue,
    linkcolor=blue,
    urlcolor=blue,
}

\renewcommand{\eqref}[1]{Eq. ({\ref{#1}})}

\begin{document}

\author{Kaitlin Gili$^{1}$, Kyle Heuton$^{1}$, Astha Shah$^{1}$, David Hammer$^{2,\Diamond}$, and Michael C. Hughes$^{1,\Diamond}$}
\affiliation{$^1$Department of Computer Science, Tufts University, Medford, MA, U.S.A.}
\affiliation{$^2$Department of Education and Department of Physics \& Astronomy, Tufts University, Medford, MA, U.S.A.}

\date{\today}

\begin{abstract}
Advances in machine learning (ML) offer new possibilities for science education research. We report on early progress in the design of an ML-based tool to analyze students’ mechanistic sensemaking, working from a coding scheme that is aligned with previous work in physics education research (PER) and amenable to recently developed ML classification strategies using language encoders. We describe pilot tests of the tool, in three versions with different language encoders, to analyze sensemaking evident in college students' written responses to brief conceptual questions. The results show, first, that the tool's measurements of sensemaking can achieve useful agreement with a human coder, and, second, that encoder design choices entail a tradeoff between accuracy and computational expense. We discuss the promise and limitations of this approach, providing examples as to how this measurement scheme may serve PER in the future. We conclude with reflections on the use of ML to support PER research, with cautious optimism for strategies of co-design between PER and ML.
\end{abstract}

\title{Combining physics education and machine learning research \\
to measure evidence of students' mechanistic sensemaking}

\maketitle

\let\thefootnote\relax\footnotetext{\noindent $\Diamond$ indicates senior authors with equal contribution}
\let\thefootnote\relax\footnotetext{\noindent Correspondence to: \url{kgili01@tufts.edu}}
\let\thefootnote\relax\footnotetext{Open access Data \& Code:~ \href{https://github.com/tufts-ml/AuSeM}{github.com/tufts-ml/AuSeM}}

\section{Introduction}

Advances in machine learning (ML) and natural language processing~\cite{BERT, touvron_llama_2023, LLM2Vec, haab2022attentioninterpretationquantitativeassessment, mylonas2022attentionmatrixdecisionfaithfulnessbased} have generated a rush to develop applications in science education, often focused on supporting the analysis and assessment of student thinking \cite{wulff2025applying}. Much of the promise is to make coding or grading of student work more efficient, as manual assessment of student thinking requires time-intensive, subjective human judgment. A key challenge is to provide 
successful ML support without burdensome requirements for expensive computational or data resources.

At the same time, scholars argue for patience and care: AI does not provide a "magic bullet" \cite{Krist2025} for research or instruction. The benefits of these new technologies will depend on careful investigation, including collaboration and co-design between ML and physics education researchers. Examples include recent work by Fussell and her colleagues to propose methodology for assessing the trustworthiness of machine coding, illustrated with analyses of students' open responses to questions about empirical science, ~\cite{PhysRevPhysEducRes.21.010128}. In subsequent work, they compare results across different large language models (LLMs) in analyzing skills evident in student lab notes~\cite{PhysRevPhysEducRes.20.010113}.   

We are focusing on the challenge of measuring the students' \emph{sensemaking}, which researchers have identified as an essential target for research and instruction, but have been constrained to study in specific episodes or small sets of data \cite{definition_sensemaking, hammer_uncertainty, vexing_questions, scientific_sensemaking, hammer_questions, sirnoorkar2023sensemakingscientificmodelingintertwined, conceptual_blends, sensemaking_identity, chemistry_sensemaking, PhysRevPhysEducRes.14.020122, Ford01072012, unpacking_sensemaking}. These studies took place in a variety of situations--different subject matter, levels of education, forms of data--and there is substantial variation across them in meanings of and nature of evidence for sensemaking. Odden and Russ~\cite{definition_sensemaking} described the literature as ``theoretically fragmented.'' To resolve this issue, they proposed a definition of sensemaking as ``a dynamic process of building or revising an explanation in order to ‘figure something out’ – to ascertain the mechanism underlying a phenomenon in order to resolve a gap or inconsistency in one’s understanding." We take this as a starting point for our work, recognizing several features we intend ultimately to study: (1) the perception of some gap or inconsistency in one's understanding; (2) reasoning with the purpose of addressing that gap; and, for science broadly and physics in particular, (3) an expectation of a \textit{mechanism} underlying the phenomenon~\cite{mech_reason}. 

For now, we focus on developing ML to analyze student reasoning with respect to mechanism, drawing on an earlier analytic framework by Russ and her colleagues for human analysis of mechanistic reasoning~\cite{mech_reason}. We review that framework below and discuss the need for its adaptation. Designing an approach to ML involves a myriad of contingencies and decisions that depend on the particularities of the situation. Here, the particular data we study consist of college students' brief 1-5 sentence explanations for answers to multiple choice conceptual questions in introductory mechanics. The data comes from from a 100-student calculus-based course at Tufts University, which used \textit{FlipIt Physics}~\cite{textbook} and emphasized students' learning to engage productively with confusion \cite{unceratinty_glue}. Given this data, two specific questions motivate our current efforts. \emph{RQ1: How can we refine an analytic framework from physics education research (PER) into a measurement scheme amenable to ML with this data set?} \emph{RQ2: For ML implementations of this scheme, how do design decisions that limit the computational resources needed impact how well the ML-provided labels agree with human coding?}

The remainder of this article is organized as follows. Sec.~\ref{Background} provides background with respect to the data we hope to analyze, prior work on mechanistic reasoning, and prior work to incorporate ML in science education research. Sec.~\ref{RQ1} presents an account of adapting the framework from Russ et al.~\cite{mech_reason} into a measurement scheme for our particular data set, addressing RQ1. Note that we will use the term \emph{scheme} and the phrase \emph{measurement scheme} throughout the paper to refer to this adaptation, which is amenable to ML. We present the scheme with examples of its application, and we describe the ML in technical detail. Part of our purpose in this account is to provide an example of convergence between PER and ML in the co-design of analytic tools. 

Sec.~\ref{deployability_analysis} delves into a specific aspect of the design, RQ2, to compare ML performance across three possible language encoders. In this section, we consider both the computational resources required for the model and the level of agreement with a human coder. We describe how this approach, which currently depends on problem-specific training, may eventually generalize to support analysis of student responses to new problems without that step of training. In Sec.~\ref{limitations}, we discuss the promises and limitations of this use of ML, specifically touching on utility, efficiency, generalizability, reliability, sustainability, and privacy. 

Finally, in Sec.~\ref{sec:conclusion}, we reflect on our efforts to date, considering both our specific aims in developing this particular tool as well as our and the broader community's more general aims of developing artificial intelligence for applications in education. As will be clear, we are not presenting settled science. Rather, we are giving examples to illustrate challenges and possibilities of this work.  

To enable others' explorations, 
we provide open-access data and code resources in a public Github repository linked at the bottom of page one, under the project name ``AuSeM''. There, we release our dataset of 385 total student responses across 4 physics problems, each manually-annotated using our approach. We also provide a codebook of instructions for human annotators, which reflects updated clarifications that emerged from our inter-rater agreement process. These resources could be used to train other models or as a guide for further manual annotation efforts of other data.  
Our provided Python code can be used to reproduce our experiments (re-train each architecture on our data) or generate mechanistic sensemaking scores for \emph{new} student responses on the same set of four problems.

\section{Background}\label{Background}

\subsection{The challenge to analyze sensemaking}\label{sensemaking_related}

The context for this effort is a calculus-based introductory course in Newtonian Mechanics at Tufts University. The course places unusual emphasis on students' sensemaking, with explicit goals of their learning to recognize and to engage proactively with confusion and uncertainty \cite{PhysRevPhysEducRes.13.020107}. The data we examine come from spring 2022, when the course was using \textit{FlipIt Physics} \cite{textbook}.
Before each class all semester long, 
students watched a 20-minute ``prelecture" video introducing new concepts, in preparation for the in-person lecture class. 
After the video, students were required to answer a multiple-choice conceptual question to check their understanding. For each problem, they submitted an answer as well as a written text explanation of their choice.

As a motivating example, consider the following problem from the fourth unit of the curriculum:
\begin{center}\label{problem4}
\textbf{Example: Problem 4} 
\fbox{\parbox{\linewidth}{
1. The net force on the box is in the positive x direction. Which of the following statements best describes the motion of the box? \\
(a) Its velocity is parallel to the x axis. \\
(b) Its acceleration is parallel to the x axis. \\
(c) Both its velocity and acceleration are parallel to the x axis. \\
(d) Neither its velocity not its acceleration need to be parallel to the x axis. \\

2. Briefly explain your answer to the previous question. 
}}
\end{center}

Our challenge is to analyze written text data from student explanations as evidence of their engagement in mechanistic sensemaking. Here are some examples of students' explanations:

\emph{E1}: ``According to Newton's second Law, F(net) = ma, we could know that the direction of a is the direction of F(net)" (Answer b)

\emph{E2}: 
``If the force is acting on it in the positive x direction, then the box must be moving in the positive x direction and will therefore have a parallel acceleration to the x axis since it is not speeding up. The velocity would also be parallel because its direction is positive. I disagree with myself as I type this but I'm not sure what my opposing argument is." (Answer c)

\emph{E3}: 
``If a 3 people were pushing the box but one person was pushing harder than the rest, the box would accelerate in the same direction as the greater force. However, with velocity, the final velocity has to take into account the initial velocity. Though the acceleration might go to the right, if the velocity was going upwards, the box might move at a slant." (Answer b)

It is a challenge to decide what to take as evidence of sensemaking, or of not-sensemaking. The first makes sense to a physicist, but is it evidence of sensemaking for the student? The second does not make sense to a physicist (the box accelerates ``since it is not speeding up"?), but the student noted experiencing uncertainty; perhaps it is evidence of their trying to identify that uncertainty. Example E3 makes sense to a physicist, and it also shows the student constructing a tangible physical situation. 

Ultimately, we hope to make progress in tapping these data for analysis of sensemaking, to study whether and how students in the course make progress in learning how to learn. In this work, we start in that direction by analyzing for evidence of mechanistic reasoning. Mechanism is consistently described in accounts of sensemaking in physics \cite{PhysRevPhysEducRes.14.020122, unpacking_sensemaking}, and there is a well-established framework for analysis \cite{mech_reason}.  

\subsection{Mechanistic reasoning in science}

Russ and her colleagues \cite{mech_reason} adapted work by philosophers of science Machamer, Darden, and Craver \cite{machamer}, who had analyzed the concept of mechanism within areas of biochemistry. This led to an analytic framework made up primarily of seven levels organized in a hierarchy: 

\#1. \textbf{Describing the target phenomenon} that is observable and reproducible.   

\#2. \textbf{Identifying setup conditions} in the spatial and temporal organization of the entities that lead to the phenomenon. 

\#3. \textbf{Identifying entities} involved in producing the phenomenon.

\#4. \textbf{Identifying activities}, the ``actions and interactions that occur among the entities." 

\#5. \textbf{Identifying properties of entities} that are relevant for producing the phenomenon.

\#6. \textbf{Identifying the spatial organization of entities} that are specifically relevant for the actions and interactions among the entities. 

\#7. \textbf{Chaining backward or forward}, in inferences about the entities and activities that could have led to (backward) or could follow from (forward) a given situation.  

These seven categories are logically hierarchical in that higher levels generally require lower levels. For example, chaining forward or backward would generally require identifying entities, properties, and activities. Russ et al.~\cite{mech_reason} added two categories they considered non-hierarchical: 

\textbf{Analogies} as explicit comparisons to other phenomena.  
\textbf{Animated models} in students' use of physical actions, whether manipulating objects available or gestures, to simulate the entities and activities they are describing. 

In the 20 years since its introduction, this framework has been applied broadly in PER and other areas of science and engineering education research \cite{org_chem_mechanistic, org_chem_ml, unpacking_sensemaking}. 

The first example explanation from our data, labeled E1 above, does not include any entities or activities; there is no evidence of mechanistic reasoning. That does not mean there was no mechanistic reasoning; only that none is evident in the provided text. Nor does it mean the reasoning is bad; it is correct.

The second explanation E2 gives clear evidence of entities (the box) and activities (its movement). Perhaps the student is also thinking of the force as an entity? Perhaps, too, there is a chaining, a force causing motion, but the logic is difficult to follow--including evidently for the student. It is ``moderate evidence of mechanistic reasoning," by Russ et al's framework. 

Example E3 has evidence of entities (people, the box), activities (pushing, motion), and of chaining (the box would accelerate in the direction of the greater force/push). There is also evidence, although it is not quite as clear, of spatial organization (if the force is sideways, by the people pushing, but the box is moving upward, the box might move at a slant). Overall, this example is ``strong evidence of mechanistic reasoning."  

Note that applying Russ et al.'s 7-level general framework to this problem involves both specification and simplification. In our data, student responses do not describe a target phenomenon, as the problem prompt does that already. Within the identification steps, there are only a small set of possibilities for what could arise as entities and activities. As we describe below, in our work to date we take this step, seeking a measurement scheme that can be applied both manually and via ML.

\subsection{Incorporating ML into science education research}\label{MLinPER_related}

ML tools are now rapidly integrating into PER, predominantly via text classification methods - algorithms that learn patterns in student text responses and sort each response into distinct categories or labels \cite{PhysRevPhysEducRes.17.020104, automated_reflections, PhysRevPhysEducRes.12.010122, PhysRevPhysEducRes.18.010141, PhysRevPhysEducRes.19.020123, PhysRevPhysEducRes.16.010142}. 
This rapid integration has been largely driven by recent advances in large language models (LLMs)~\citep{BERT,touvron_llama_2023, jiang2023mistral7b}
trained on vast amounts of language corpora (books, Wikipedia, websites, etc). 
The availability of chatbots like ChatGPT powered by LLMs makes it possible for PER researchers to directly feed text into a chatbot window to be classified~\cite{ChatGPT_Ed}. Below, we contrast different strategies for using LLMs in PER.

One paradigm for LLM-enabled PER is \emph{in-context learning} (ICL), 
where an existing chatbot is provided a student text response and prompted via natural language instructions to classify it into a desired labeling system.
To be clear, the ``learning" here is euphemistic; ICL does not entail retraining the LLM itself at all.
Instead, it relies on the pretrained ability of the LLM to transform an instructional prompt and provided input text into useful output text.
ICL is data-efficient, with proponents often providing only a few example demonstrations in the instructions to clarify the desired labeling transformation.
However, ICL has a number of important limitations. 
First, ICL's utility relies on the assumption that the LLM is useful for a specific downstream language task in education (correctness, reasoning, sensemaking, etc.). 
The underlying LLM was not trained to do well on these tasks, and it is important to verify at scale whether an LLM is reliable at certain tasks. 
ML researchers have sometimes questioned the utility of ICL, showing that ``demonstrations have a marginal impact''~\citep{longDoesInContextLearning2024a} and ``randomly replacing labels in the demonstrations barely hurts performance''~\citep{minRethinkingRoleDemonstrations2022}.
Second, ICL is sensitive and variable. Outputs can vary depending on the exact prompt~\citep{jiangHowCanWe2020} and the selection and order of examples~\citep{liuWhatMakesGood2022}. Even the same exact input may lead to different outputs due to the stochastic nature of chatbots.
Third, LLMs reflect biases inherent in the source data~\citep{liangUnderstandingMitigatingSocial2021}.
Finally, there is a lack of transparency: closed-source versions such as ChatGPT do not allow users to inspect, understand, or modify how outputs are obtained.
For these reason, in-context learning methods deserve and require further investigation for their reliability in science education research. 

Other works focus on training supervised ML predictors to accomplish their specific task, integrating accepted human annotation frameworks in science education with machine learning practices \cite{PhysRevPhysEducRes.12.010122, PhysRevPhysEducRes.18.010141, org_chem_ml}. Watts et al.~\cite{org_chem_ml} is a prime example of this type of integration between science education and machine learning research. Watts et al.~\cite{org_chem_mechanistic} had previously adapted Russ et al.~\cite{mech_reason} for use in organic chemistry; follow-up work ~\cite{org_chem_ml} adapts further to apply ML binary classifiers. This entailed separating Russ's original categories into stand-alone binary (present or not present) codes specific to organic chemistry problems. For example, they defined the specific code \emph{electron movement} within Russ's \emph{identifying activities}. The authors treat each code as independent evidence of mechanistic reasoning, without the hierarchal organization feature of Russ's framework, which simplifies the ML methodology, a useful first  step towards integrating ML and science education research. Student responses are used as training data for separate binary classifiers for each code, where various classifier types are attempted, and only the final results of the highest performing classifier are reported. 

Similar to Watts et al, we take the step of adapting the framework to specific features of the tasks in our data set, and we simplify to binary codes, as we describe in the following section. We will discuss similarities and differences in technical detail in Sec.~\ref{model_frameworks}.

Fussell et al.~\cite{PhysRevPhysEducRes.21.010128} provide an evaluation of different large-language models for measuring specific skills in students’ typed lab notes. The authors compare off-the-shelf pre-trained open-source LLMs with ones that are \emph{fine-tuned} (further trained) on training data specific to their task. 
The results showed that higher-resource models (e.g. more training data, computational power) often outperform lower-resource models for their particular task, although depending on the evaluation metric there are exceptions. More specifically, the authors found that all fine-tuned models outperformed the off-the-shelf pre-trained models, and they suggest that for theory-driven classification tasks (i.e. where there is a specific task in science education), fine-tuning might be more useful. 

There are a few key distinctions between our current work and Fussell et al. Similarly to Fussell et al., we compare the performances of ML model variants in the context of our specific task; however, we do not attempt to provide more general insights about the models on other tasks or datasets. We will discuss the similarities and differences between Fussell et al. and our work in more technical detail in 
in Sec.~\ref{model_frameworks}.

\section{RQ1: PER-ML Measurement Scheme}\label{RQ1}

\subsection{Refining the framework for a particular data set}\label{8_domains}

Russ et al.~\cite{mech_reason} designed their framework to be applicable across areas and levels of scientific reasoning. That breadth of applicability has been helpful for the research community, but its generality presents challenges for the development of an ML tool with context-specific datasets. Given a particular data set, such as the conceptual question above about a net force acting on a box, it is possible and advantageous to tune it to the specifics of the task. The strengths and limitations of machine learning motivate other simplifications as well.    

\textbf{Task-specific considerations.} We are interested in measuring student responses to how they solved a \emph{particular} problem within a set of closed-ended introductory physics problems. Each problem contains its own specific entities, properties, activities, etc. that are relevant. The box question, for example, mentions only one entity, and there is a limited set of entities (such as people to push) that students might add to the situation. As such, the mechanistic evidence that we are looking for is problem-dependent, which allows us to simplify the scheme in several ways. 

\emph{(1) No need to re-describe the target phenomenon}: 
Each problem's instruction text makes the target phenomenon clear; students focused on \emph{brief} explanations and thus did not restate that phenomenon in their own words. For these reasons, we do not include this first code from Russ's framework in our scheme. 

\emph{(2) Restricted set of entities, properties, and activities}: The ``net force on the box'' example problem names only the box as an entity. There are in principle other entities one could mention as involved, such as the Earth or the surface under the box, but no students in this data set mention them. For each problem in our data, we confine our analysis to a limited, analyst-defined set of entities, properties, and activities that could reasonably be relevant or that arise in the data.  

\textbf{ML design considerations.} While we want custom annotations for each problem, developing a separate classifier for each problem would be prohibitively expensive and unlikely to work well without many training examples.
We are thus interested in combining powerful language encoders \cite{BERT,LLM2Vec, jiang2023mistral7b, touvron_llama_2023} with \emph{shared} classification heads that take advantage of similar word occurrence and grammatical structure across problems to \emph{efficiently} annotate new examples. As such, it is important to construct multi-label classification categories that share similar linguistic features across student responses and problems. This influences three key design decisions. 

\emph{(1) Share initial and subsequent spatial organizations}: We assume a description of the spatial organization  of entities would contain similar semantic and structural features independent of whether it refers to the initial, intermediate, or final conditions. These temporal states can be represented by a separate context embedding (further discussed in Sec.~\ref{model_frameworks}), such that evidence for each temporal state can still be individually annotated, but spatial linguistic features can be shared.

\emph{(2) Separate central from non-central entities}: By definition, central entities are connected to the other codes (properties, activities, etc.). As such, central and non-central entities will contain dissimilar relationships among words in a student response. Keeping these categories separate is especially useful for future work that incorporates dependencies between domain labels.

\emph{(3) Separate movement activities from interactions}: Descriptions of individual entity movement activities (e.g. the object is moving with an initial velocity) are different from descriptions of the interactions between entities (e.g. the hand pushes the book) both semantically and structurally within a sentence. 

Like Watts et al.~\cite{org_chem_ml}, we relax the hierarchy of Russ's framework, coding only for the presence or absence of evidence in each category. This simplifies our annotation scheme and ML method for now; later in Sec.~\ref{sec:conclusion} we discuss plans going forward to reintroduce some form of that hierarchy. This would be an improvement on our current modeling assumptions: for example, evidence of properties is likely to correlate with evidence of entities.

\begin{figure*}
{    
    \includegraphics[width = \linewidth]{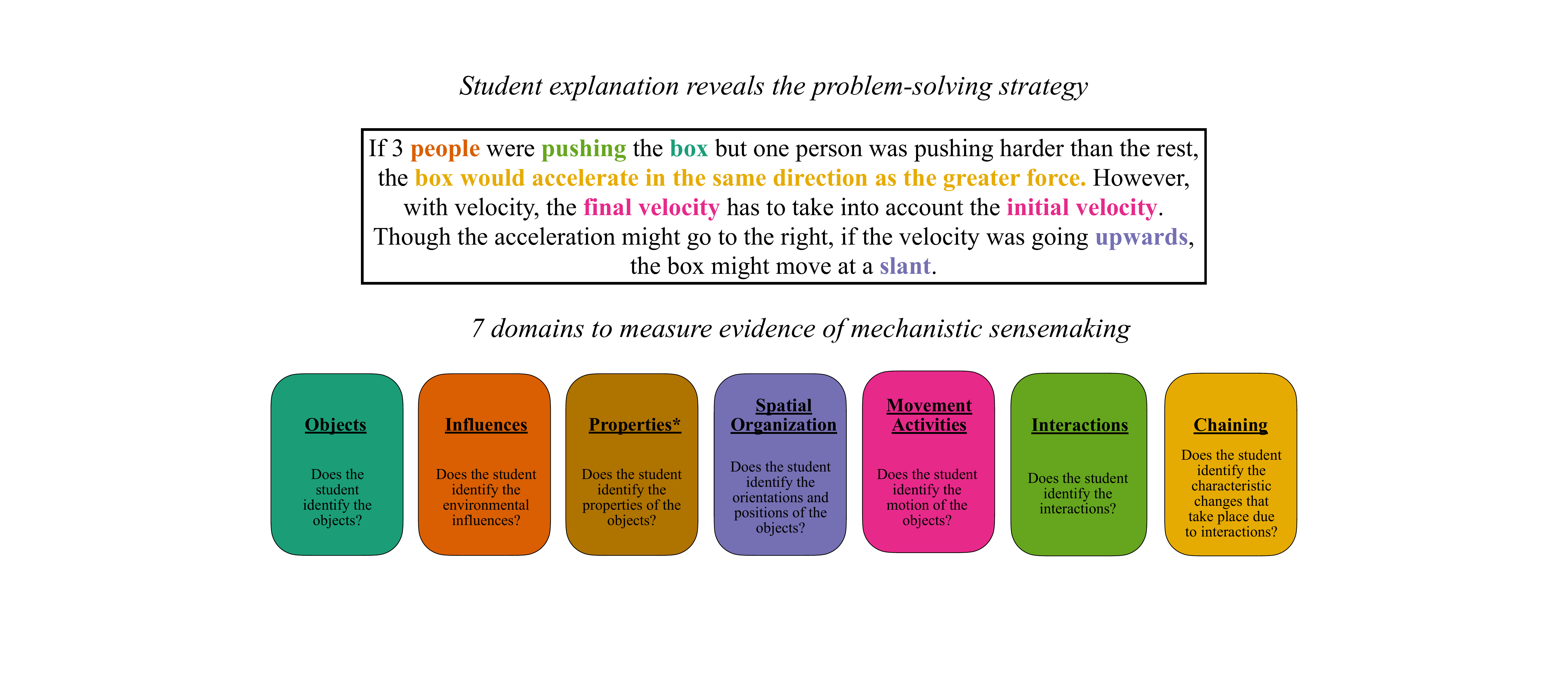}
}
    \caption{\textbf{An annotated student text explanation to a ``force on a box'' problem (defined in Sec.~\ref{problem4}).} We display the 7 domains used to measure evidence of a mechanistic sensemaking process in the student's text. Each domain may contain a different number of yes/no criteria questions to answer dependent on the problem. We color words and phrases that would prompt yes answers to the criteria questions within that same-colored domain. The * next to a domain name signifies that the student did not explicitly identify its criteria. More than one word or phrase can be annotated the same color if  multiplicities exist for that domain (e.g. spatial organization  1: ``upwards'' , spatial organization  2: ``slant''). This particular student displays \emph{high evidence} of mechanistic sensemaking ($s = 0.89$) by imagining a scenario where multiple interactions influence the box's dynamics and describing the mechanistic relationship between a net external force and a change in the box's motion.}
    \label{fig1}
\end{figure*}

\textbf{Proposed domains.}
Given the above considerations, we propose 7 domains, broad categories of yes/no questions that measure evidence of mechanistic sensemaking in student thinking via the framework:

\begin{enumerate}[leftmargin=*,noitemsep]

\item \textbf{Objects.}~Identifying the central entities that interact to comprise the physical phenomena in a single hierarchical level. These objects have properties, spatial layout, and movement relevant to the solution process. Examples: ``car'', ``box'', ``ball'', ``particles'', etc.

\item \textbf{Influences.}~Identifying the non-central environmental entities that affect the objects and their activities. Only the influence $\rightarrow$ object interaction is relevant to the phenomena. Examples: ``Earth's gravitational field'', ``human hand'', ``electromagnetic field''. 

\item \textbf{Properties.}~Identifying the relevant qualities that are confined to the physical boundary of an object. Examples: mass, temperature, volume.

\item \textbf{Spatial organization.}~Identifying the spatial states or orientations of an object. Examples: ``downwards'', ``upwards'', ``left'', ``right'', ``x-direction'', ``y-direction''.

\item \textbf{Movement activities.}~Identifying the motion states of an object. Examples: ``initial velocity'', ``slowing'', ``speeding'', ``accelerating''.

\item \textbf{Interactions.}~Identifying the physical exchanges between all object $\leftarrow\rightarrow$ object and influence $\rightarrow$ object pairs. Examples: ``gravitational force'', ``human pushing'', ``electromagnetic force''.

\item \textbf{Chaining.}~Identifying the characteristic (a property, spatial organization, or movement activity) changes that result from specific object $\leftarrow\rightarrow$ object interactions and influence $\rightarrow$ object interactions. Examples: ``constant acceleration of a ball caused by Earth's gravitational field exerting a downward net force''; ``acceleration of a book initially at rest caused by a human push''.
\end{enumerate}

Our scheme corresponds with Russ's framework as follows: the properties domain directly maps to Russ's category \#5; the objects domain and influences domain combine to obtain \#3; the spatial organization domain encompasses \#2 and \#6; the movement activities and interactions domains combine to form \#4, and the chaining domain maps to \#7. We do not consider stating the target phenomena, analogies, or mental model animation in this work - although these are interesting additions for future research.

\textbf{Example problem and associated coding.}
Consider again the ``force on a box'' problem from Sec.~\ref{problem4}. An example student explanation is shown in Fig.~\ref{fig1}, with specific parts color-coded by our 7 domains. This coding demonstrates how parts of an individual student response can be associated to each of the 7 domains. These specific parts could be later used to answer yes/no questions to quantify evidence of sensemaking, as described in Sec.~\ref{binary_annotation}.
Some domains correspond to multiple parts of the response (e.g. the object has more than one spatial organization in the context of the overall problem).
Note that not all 7 domains need to be coded for if some are irrelevant to a problem, though all domains are relevant to the example problem here. 

In sub-sections~\ref{binary_annotation} and \ref{model_frameworks} below, we introduce how we apply quantitative labels to measure mechanistic sensemaking via our scheme and operationalize our modular classification model, for readers who are interested in the technical details.

\subsection{Measurement as multi-label binary annotation}\label{binary_annotation}

Given a specific physics problem, our measurement scheme requires the annotator to develop a list of yes/no \emph{criteria questions} for each of the 7 domains that is problem-specific. Thus, given a dataset of potential problems and student responses to be measured, an annotator can follow the proposed steps: 

\textbf{Step 1.}~Select $P$ problems from an introductory physics course on classical mechanics, where there exists a physical target phenomena to be understood. 

\textbf{Step 2.} For each problem $p \in \{1,..,P\}$ and for each domain $d \in \{1,..,7\}$, determine the set of yes/no criteria questions $C^p_{d}$ that are pertinent to the specific problem. For example, in Problem 4 from Sec.~\ref{problem4}, there is only a single object, and thus $|C^4_{1}| = 1$ with the criteria question \emph{Does the student identify the object?}. However, there are two spatial organization criteria questions $|C^4_{4}| = 2$: \emph{Does the student identify the object spatial organization before the net applied force?} \emph{Does the student identify the object spatial organization after the net applied force?} We denote the total number of criteria questions for problem $p$ as $ T_{p} = \sum_{d=1}^{7}| C^p_{d}|$. The full list of criteria for Problem 4 is displayed in the Appendix \ref{dataset_details}.

\textbf{Step 3.}~Given a student's text response to problem $p$, the annotator will use the yes/no criteria questions to obtain a vector of binary digits representing their answers. The resulting criteria vector for a student response $n$ is denoted as $\textbf{y}^{n} \in \{0,1\}^{T_p}$. To further compute a sensemaking score $s \in [0,1]$ one can sum the number of $1s$ and divide by the length of the vector $T_{p}$. This score is to be interpreted as the \emph{amount of evidence of mechanistic sensemaking}. An annotator could decide to weight the domains (or specific criteria) differently. In this work, we keep the weights equal. 
To finish our example, the criteria vector for the student response in Fig.~\ref{fig1} is ${\textbf{y}}^{n} = [1,1,0,1,1,1,1,1,1]$ with the corresponding sensemaking score $0.89$, a ``yes" code for all domains but ``property." Considering the response itself, a high score seems appropriate: The student imagined a tangible scenario with multiple influences on the box.  

\subsection{Multi-label binary classification model}\label{model_frameworks}


Here, we describe an ML model we constructed for our task, that can automatically map a student text response to a numerical sensemaking score, where a higher value indicates greater evidence of mechanistic sensemaking. While our approach relies on problem-specific criteria, some aspects of the model are shared across problems to gain statistical strength.
The overall model is modular: components like the encoder can be varied to tradeoff between model capacity and resource consumption.

\textbf{From text to embeddings.} Given a student response $n$, we can select an \emph{encoder} to map all text in the response to a numerical representation $\textbf{x}^n$. A simple encoder design is \emph{counting}; we determine a vocabulary $\mathcal{V}$ as the set of words across all student responses in the training dataset, with an additional vocabulary entry for unknown words, and record the number of times each one appears in the $n^{th}$ example. This numerical representation, or embedding $\textbf{x}^{n} \in \mathbb{R}^{|\mathcal{V}|}$, typically called \emph{Bag-of-Words}, reduces the richness of the text explanation to the pattern of word occurrence \cite{bag_of_words,pml1Book}. From here on, we will refer to this encoder as \texttt{BoW}.    

\begin{figure*}
    \centering
    \includegraphics[width = \linewidth]{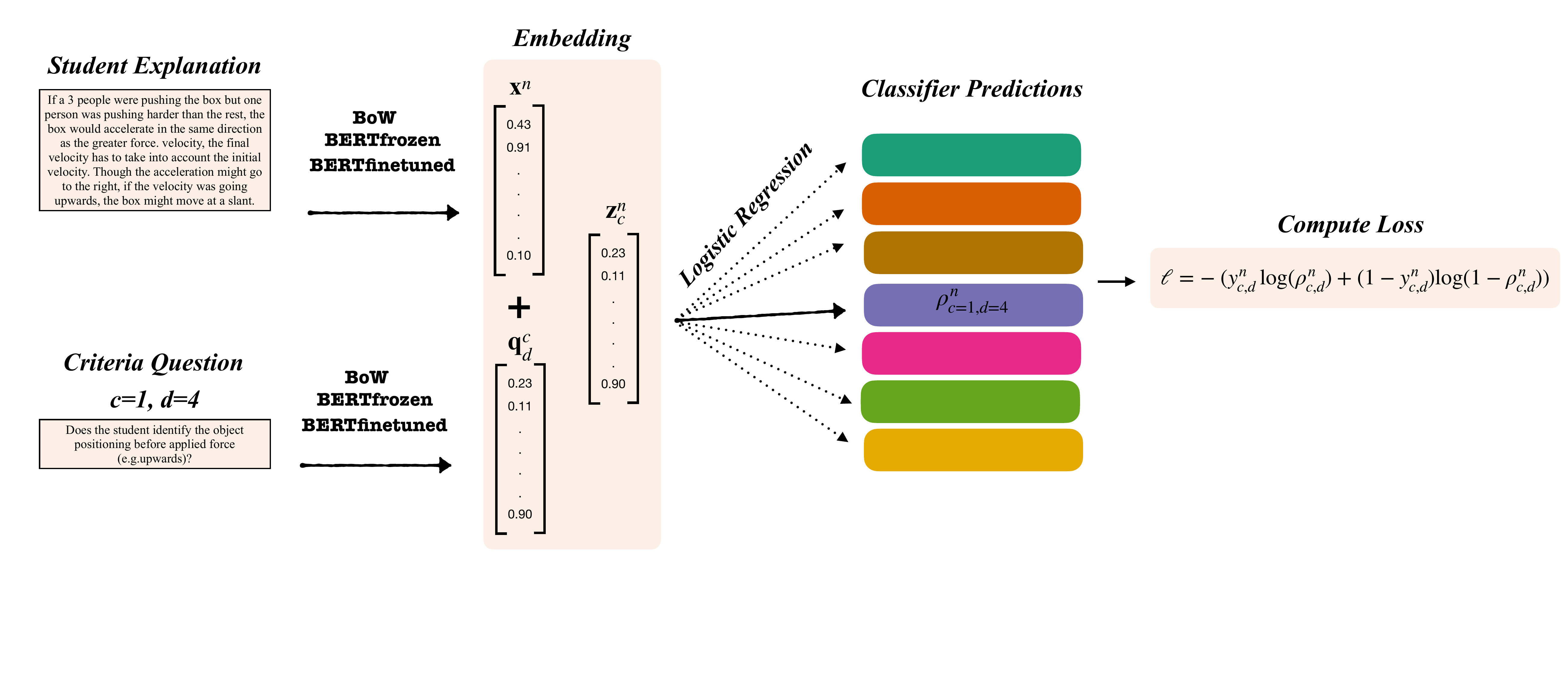}
    \caption{\textbf{How the proposed model makes a probabilistic prediction to answer a yes/no criteria question given a student response.} Using a selected encoder (\texttt{BoW}, \texttt{BERTfrozen}, \texttt{BERTfinetuned}), the student response is embedded into a numerical vector representation ${\textbf{x}}^{n}$. An embedded representation of the criteria context ${\textbf{q}}^{c}$ is added to ${\textbf{x}}^{n}$, producing an updated embedding ${\textbf{z}}_{c}^{n}$ contains information about both the student response $n$ and the  current criteria question $c$.
    Finally, out of 7 possible domain-specific logistic regression classifiers, the current question's domain $d$ selects its specific classifier to produce a predicted probability $\rho_{c,d}^n$ of answering ``yes'' to the criteria question.
    As a training objective, we use the cross-entropy loss comparing this probability to supervisory ground-truth label $y^{n}_{c,d}$. Each LR classifier's weights and bias are updated via gradient-descent to minimize the sum of this loss over the entire training set. When using \texttt{BERTfinetuned} as the encoder, encoder parameters are also updated to minimize the loss.}
    \label{fig2}
\end{figure*}

To capture more complex patterns (e.g relative co-occurrence between all words in the response, in-context word occurrence, and word positional associations with grammar rules), one can use a \emph{large language model (LLM) encoder} \cite{turner2024}. For example, Google's \emph{BERT} (Bi-directional Encoder Representations from Transformers) model combines tokenization, self-attention, and a multi-layer perceptron neural network - all pre-trained on English Wikipedia and BookCorpus to generate token embeddings \cite{BERT}. A token is a set of characters containing a unique ID in BERT's vocabulary, such that an entire text can be converted to tokens with individual embeddings $\textbf{x}^{n}_{token}\in \mathbb{R}^{768}$, which carry information about all other tokens in the passage. BERT also uses a sentence-level embedding, known as $\textbf{x}^{n}_{CLS} \in \mathbb{R}^{768}$, which can be used as a full-response representation rather than using distinct representations for each token. From here on, we will refer to this encoder as \texttt{BERTfrozen}. As a final encoder option, the sentence embedding $\textbf{x}^{n}_{CLS}$ can be fine-tuned by training BERT's previously trained weights $\{\boldsymbol\theta\}$ on new downstream classification tasks \cite{sun2020finetunebert}. We will refer to this encoder as \texttt{BERTfinetuned}. This allows for the $\textbf{x}^{n}_{CLS}$ embedding to capture patterns in the text that are specific to the target prediction task. It requires significantly more computational resources, however. Current LLMs, such as Google's Gemini or OpenAI's ChatGPT, require far more still. Part of our interest is to minimize the computational demands of the ML.  


\textbf{Shared classifiers across problems.} Given an embedding representation $\textbf{x}^{n}$ for a student response, we want to correctly assign it to a criteria vector $\textbf{y}^n$. Thus, we have a multi-label task \cite{multi_label_overview}. If we were to construct \emph{separate} binary classification models for each criteria question, then we would be required to train $T_{p} * P$ independent classifiers. Rather, we propose to work off of an informed assumption: shared domains will have similar word occurrence and grammatical structure across $P$ problems. For example, the word “speeding” should indicate evidence of sensemaking in the movement activities domain regardless of the problem. Furthermore, words that should be classified as objects will share similar positions in the sentence across problems due to object-subject-verb rules. Thus, for $P$ problems, we propose to train 7 classifier heads, one for each domain. We reduce from a scaling linear in $P$ to constant in $P$. 

In order for this shared set of  classifiers to effectively annotate multiple problems each with multiple yes/no criteria questions, 
we design each classifier to take an additional input beyond the embedding of student response text: a criteria \emph{context} embedding $\textbf{q}^c_d$ that encodes the text of the yes/no question being asked (e.g. \emph{Does the student identify the object spatial organization before applied force (e.g. upwards)?}). The index $c \in \{1,..,|C^{p}_{d}|\}$ specifies the criteria question the classifier should label. 

Each classifier must then take in student response embedding $\mathbf{x}^n$ and criteria context embedding $\mathbf{q}^c_d$. 
There are multiple possible ways to combine the two inputs into a joint representation $\textbf{z}^n_{c}$ of response and context vectors that is the ultimate feature vector for the classifier.
As a starting point, we have selected a simple approach of adding the embeddings together: $\textbf{z}^n_{c,d} = \textbf{q}^c_d + \textbf{x}^n$. 
Possible alternatives would be to concatenate the two vectors together, or to produce one embedding from the combined text of question and response. We leave the exploration of alternative methods for future work. 

Fig.~\ref{fig2} visually illustrates how the model makes a probabilistic prediction for a specific yes/no criterion given a student's response to a question. Given embedding $\textbf{z}^n_{c,d}$, we use a domain-specific logistic regression (LR) classifier to produce the predicted probability $\rho^n_{c,d} \in (0,1)$. We set $\rho^n_{c,d} = \sigma( \textbf{W}_d \textbf{z}^n_{c,d} + b_d)$, where $\textbf{W}_d, b_d$ are the weight vector and bias scalar for domain $d$, and $\sigma$ denotes the logistic sigmoid function.

The figure also shows how that prediction impacts the loss, which measures the difference between the model's current predictions and human coding. For training, we have a dataset of student embeddings with human annotated labels (the criteria vectors) $D_{\text{Train}} =\{\textbf{x}^n, \textbf{y}^n\}_{n=1}^{N}$.
Each $y^n_{c,d} \in \{0,1\}$ represents a single binary label with respect to criteria question $c$ of domain $d$ for the student response $n$.
Each $y^n_{c,d}$ is assumed conditionally independent of other labels given $\textbf{x}^n$ and all model parameters.
We minimize the following binary cross-entropy loss
\begin{equation}
\ell(\rho^n_{c,d},y^n_{c,d}) = -y^n_{c,d}\log(\rho^n_{c,d}) - (1-y^n_{c,d})\log(1-\rho^n_{c,d}).
   \tag{1}
\end{equation}
This loss is \emph{aggregated} over all explanations $n$, domains $d$, and criteria $c$.
We also include an added regularization term that penalizes the sum-of-squares of the weight parameters.
For the precise loss notation utilized in our methods including regularization terms, see Appendix \ref{detailed_loss}. The training procedure encourages all 7 classifiers to accurately predict their assigned pieces of the criteria vector $\textbf{y}^n$ for an embedded student response $\textbf{x}^{n}$. 

For some encoder choices, we do not update any encoder parameters during training; only the classifier weights and biases $\{\textbf{W}_d, b_{d}\}_{d=1}^D$ are updated via gradient descent. When \emph{fine-tuning} BERT, our gradient descent procedure updates encoder parameters $\{\boldsymbol\theta\}$ alongside classifier weights. Both are explored in our model comparative analysis in  Sec.~\ref{deployability_analysis}. 

\textbf{Technical comparison with related work.} Similar to Fussell et al.~\cite{PhysRevPhysEducRes.21.010128}, we train a \emph{BagofWords} (\texttt{BoW}) encoder and a \texttt{BERTfinetuned} encoder. We also observed that the \texttt{BERTfinetuned} encoder achieves better performance than the \texttt{BoW} encoder for our specific task. In general, our later results support their conclusion that fine-tuning adds value for a theory-driven classification in PER. They also compared two model architectures with multiple LLaMa encoders~\cite{touvron_llama_2023}, which require significantly more computational resources than our BERT encoders.  Similar to Watts et al.~\cite{org_chem_ml}, we train multiple binary classifiers for task-specific evidence of mechanistic reasoning. They determined convolutional neural networks (CNNs) to outperform other model architectures such as naive Bayes, logistic regression, support vector machines, and transformers. Their results show high accuracy on the test data with slightly lower Cohen's Kappa values to account for the data imbalance. We chose to continue with BERT encoders, expecting better inductive bias for our task than CNNs, especially when pretrained on large language corpora then fine-tuned on our limited data.

\section{RQ2: How ML Design Choices Can Impact Performance}\label{deployability_analysis}

We seek a performance comparison of the proposed classification models outlined in Sec.~\ref{model_frameworks} with different encoding and training strategies. More specifically, our intention is to estimate how these models will perform given explanations from new students that it has never seen before, but on the same set of problems. Below, we outline our implementation and evaluation procedures, then present and discuss results.

\begin{figure*}[ht]
  \centering
  \begin{subfigure}[b]{0.9\textwidth}\centering
\includegraphics[width=\linewidth]{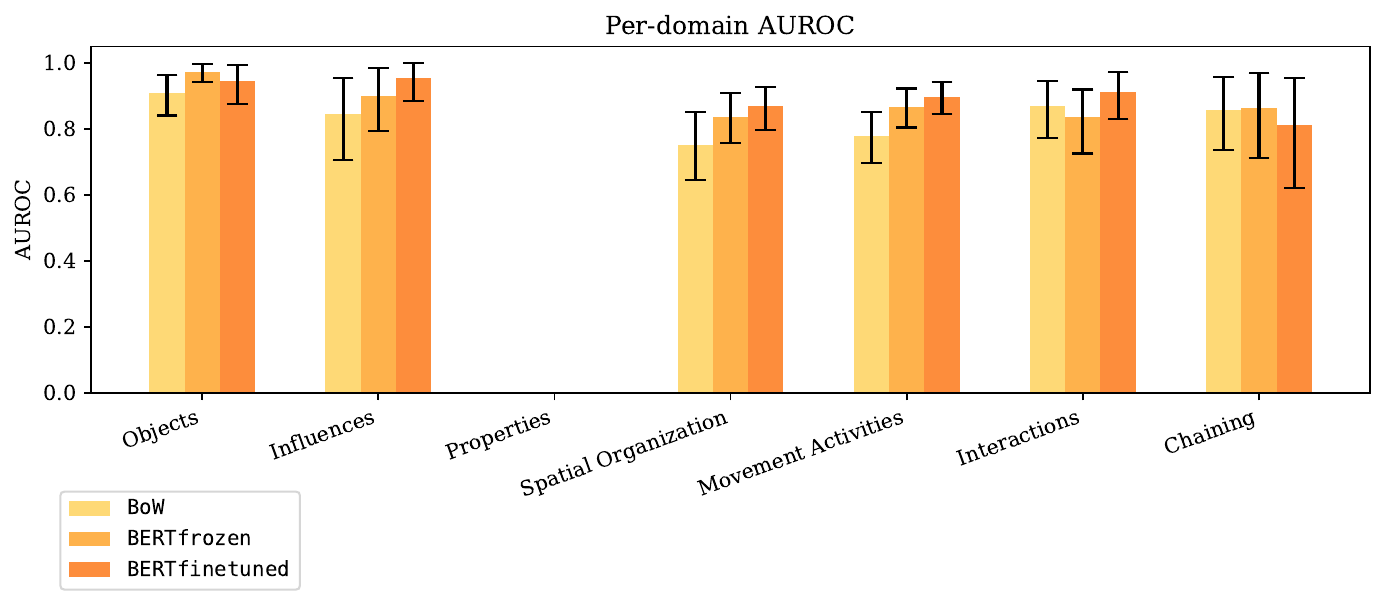}
  \end{subfigure}
  \caption{\textbf{Model performance across sensemaking domains.} For each domain, we show the average AUROC and 95\% confidence interval from our 1000 bootstrapped test samples. Here, we see that \texttt{BERTfrozen} has the best performance in 2 domains (objects and chaining), and \texttt{BERTfinetuned} has a higher average AUROC in the other 5 (other than properties). We observe that all 3 models are quite good at detecting evidence that a student has sensemade an object, but do worse on domains with less data. The properties domain only has 8 positive examples in the entire dataset, which were all used in training. As such, we are unable to report the AUROC for the properties domain.}
  \label{domain_bar}
\end{figure*}

\textbf{Dataset of student responses.} Our dataset consists of 385 total de-identified student-responses across four problems distributed to 100 students in the first few lectures of the 2022 Tufts University Introduction to Physics course. The retrospective analysis of this deidentified human subjects data was approved by our institutional review board. 

All problems have the same structure as Problem 4 in Sec.~\ref{problem4} shown above and are presented to the students as pre-lecture digital assignments \cite{textbook}. 
See Appendix \ref{dataset_details} for exact problem text.
The same $100$ students are asked to solve each problem and explain how they obtained their solution. A few students chose not to respond to a particular problem, which brings our total number of explanations to $385$. 

\textbf{Annotating ground-truth labels.}
Complete annotation for all 4 problems was carried out by a domain expert: the first author of this work who holds a doctorate in physics. This included defining the yes/no criteria questions for each problem (provided in Appendix \ref{dataset_details}) and assigning binary values for each student response.
These problems vary in criteria vector size: $\{T_{1} = 7, \ T_{2} = 11, \ T_3 = 11, \ T_4 = 9\}$.
Some descriptive statistics summarizing annotations across problems are provided in Appendix \ref{annotation_details}.
We validated inter-rater reliability post-hoc, as described in Sec.~\ref{sec:reliability}.
In future work, we plan to incorporate training examples from multiple expert annotators. 

\textbf{Experimental setup.} To evaluate model performance on never-before-seen responses, we randomly select 15 students who responded to all four questions to use as a held-out test set. We use the rest of the data (325 samples) for training and validation. To prevent data leakage, we stratify the data by student to ensure that all responses from the same student across 4 problems end up either exclusively in the training set or exclusively in the test set.

Using this train/test split, we fit the models \texttt{BoW}, \texttt{BERTfrozen}, and \texttt{BERTfinetuned} described earlier.
All models are trained with the Adam optimizer \cite{kingma_adam_2015} for $1000$ epochs. Convergence was verified by observing loss function behavior. We perform a grid search over key hyperparameters: learning rate $\alpha$ and regularization strength $\lambda$ ($\lambda_1$ for classifier weights and $\lambda_2$ for \emph{BERT} model weights); reproducible details are available in Appendix \ref{hyper-param-procedure}.
For \texttt{BoW} and \texttt{BERTfrozen}, we use 5-fold cross-validation, also stratified by student, to select hyperparameters. We pick the combination with the best average AUROC score across folds. 

The primary performance metric we report and evaluate is the \emph{AUROC}, the Area Under the Receiver Operating Characteristic (ROC) curve. For a given classifier, this curve depicts the true positive rates (y-axis) and false positive rates (x-axis) achieved across different numerical thresholds for turning predicted probabilities into binary decisions. There are two main reasons we believe AUROC to be the most useful metric for our context: First, AUROC does not require us to choose an arbitrary probability threshold to distinguish label 1 predictions from label 0 predictions. Instead AUROC sums over all possible thresholds, providing an overall numerical summary where increasing values indicate better performance. 
AUROC is bounded in $(0,1)$.
Second, due to imbalanced annotations in our training dataset (i.e. some multi-label tasks have more of the positive or negative class rather than a 50-50 split), AUROC is an easier to interpret metric to report than precision or accuracy, as its chance baseline is $0.5$ independent of imbalance. For these reasons, we use AUROC as the primary performance metric throughout our analyses.

For the interested reader, we also include additional metrics like area under the precision-recall curve in the Appendix \ref{auprc_metrics} and confusion matrices in Appendix \ref{CONFUSION}. 

\textbf{Fine-tuning details.}
For \texttt{BERTfinetuned}, we require a larger amount of computational resources per experiment.
Our grid search over hyperparameters entails 324 separate efforts to train this encoder's millions of weights.
We thus used a single training and validation set to run the hyperparameter grid search, avoiding the extra costs of 5-fold cross-validation. 
To better preserve the prior information learned in \emph{BERT's} original training, we use L2SP regularization \cite{chelba2006adaptation, xuhong2018explicit} when training \texttt{BERTfinetuned}.
This means we enforce two separate sum-of-squared-difference penalties in the loss, one for \emph{BERT} encoder weights and another for the LR classifier head weights. 
Each has a different desired mean and a different strength hyperparameter that is tuned.
We set the batch-size to $N_{batch}=32$, which has been shown empirically to work well with deep neural networks \cite{masters2018revisitingsmallbatchtraining}.~

During finetuning, we originally incorporated an $8^{th}$ domain classifier that we now believe to be out of scope for this particular work. As the domain-specific classifiers are independent, this does not influence the results of \texttt{BoW} or \texttt{BERTfrozen}. However, we acknowledge this may impact our reported performance of \texttt{BERTfinetuned} as the embeddings across classifiers are shared. 
We chose not to retrain as the model resources required to do another hyperparameter optimization (another set of 324 HPC experiments) is high. Re-training without the $8^{th}$ domain would likely only increase the performance of \texttt{BERTfinetuned} on 7 key domains, and as we show below it is already performing better than other variants . 

\textbf{Results from initial validation phase.}
The best model for \texttt{BoW} achieved an average AUROC score of $0.839$ across validation folds with the hyper-parameters:~$\alpha=0.01, \lambda_{1}=0.001$.~Similarly, for \texttt{BERTfrozen}, the best model achieved an AUROC score of $0.890$ with the hyper-parameters:~$\alpha = 0.001, \lambda_{1} = 0.001$.  
For \texttt{BERTfinetuned}, after running 324 fine-tuning experiments on Tufts' HPC services, we picked a model with final AUROC of $0.916$ on the validation set, which had the following hyperparameters: $\alpha = 1e-5$, $\lambda_1 = 0.0001$, and $\lambda_2 = 0.1$.

\textbf{Results on heldout test data.} We retrain our selected models (with model-specific optimal hyperparameter settings) on all training and validation data. We then evaluate on the held-out test set of 60 responses from $15$ randomly selected students. In Fig.~\ref{domain_bar}, we show the average AUROC across independent test predictions for each of the 7 domain classifiers and for each model type. 
To create approximate uncertainty intervals, we use bootstrap resampling of our test set~\cite{Efron1992, raschka2020modelevaluationmodelselection} to generate $1000$ same-sized test sets that represent plausible unseen student responses, and compute the AUROC of each one. Intervals in Fig.~\ref{domain_bar} depict the 2.5-97.5th percentile range of these bootstrapped AUROCs.

Beyond the per-domain results in Fig.~\ref{domain_bar}, we show the overall average AUROC for each model in Table \ref{table2}. This micro-average is computed by weighting each annotation label equally across all domains. As such, these values represent the annotation quality of the multi-task classification set-up as a whole for each model. In Table \ref{table2}, we also report the time taken to train the final selected model. To assess the significance of differences between models, we provide the $95\%$ confidence interval in Table \ref{table2} in the \emph{difference} in AUROC between models, computed by taking percentiles from our bootstrapping procedure.
These intervals can be interpreted as indicating uncertainty in AUROC difference across models. When comparing model A and B, if the interval is all positive and an exact zero is outside of the interval, we can be $95\%$ confident that there exists an actual positive difference between models, and model A is indeed better than B. 

\texttt{BoW} performs decently overall with an AUROC score of $0.839$ and trains efficiently in 1 minute. For the objects domain in particular, it performs nearly as well as the other two models. 
\texttt{BERTfrozen} performs slightly better with an AUROC score of $0.887$ and takes 7 minutes to train. It performs the best for two domains: objects and chaining. It is able to reliably outperform \texttt{BoW} on average, as zero in not within the confidence interval for AUROC difference reported in Table \ref{table2}. (The interval containing zero would indicate that no significant difference between models is plausible.)  
\texttt{BERTfinetuned} achieves an AUROC score of $0.916$. It performs the best for influences, spatial organization, movement activities, and interactions.  It takes significant time to train, 160 minutes. That time might not be prohibitive, depending on the resources available; training is only needed when new student data is available. 

One major limitation in the results is that we cannot report AUROC results for yes/no criteria labels in the properties domain in Fig.~\ref{domain_bar}. In our dataset, class imbalance is extremely large for this domain, and no positive examples were included in the test set, making AUROC undefined.
More data, especially more positive examples, are necessary for improvement. 

\begin{table}[ht]
\centering
{ 
\begin{tabular}{lcc}
    \toprule
    & \textbf{AUROC (95\% CI)} & \textbf{Train time} \\
    \midrule
    \textbf{\texttt{BoW}} & 0.839 (0.803, 0.873) & 1 min \\
    \textbf{\texttt{BERTfrozen}}  & 0.887 (0.860, 0.913) & 7 min \\
    \textbf{\texttt{BERTfinetuned}} & 0.916 (0.893, 0.938) & 160 min \\
    \bottomrule
\end{tabular}

\vspace{1em} 

\begin{tabular}{lc}
    \toprule
    & \textbf{Avg. (95\% CI)} \\
    \midrule
    $\Delta \text{AUROC}_{\texttt{Finetuned} - \texttt{Frozen}}$ & 0.029  (0.009, 0.049) \\
    $\Delta \text{AUROC}_{\texttt{Finetuned} - \texttt{BoW}}$ & 0.077  (0.052, 0.104) \\
    $\Delta  \text{AUROC}_{\texttt{Frozen} - \texttt{BoW}}$ &  0.048 (0.021, 0.075) \\
    \bottomrule
\end{tabular}
}
\caption{Avg.~model performance and comparison.}
\label{table2}
\end{table}

\section{Benefits and Limitations}\label{limitations}

\subsection{Utility.}~As we described in the introduction, we are working to automate the measurement of mechanistic sensemaking \emph{evidence} for research purposes across a set of student responses to conceptual problems in introductory Newtonian mechanics. This will allow larger-scale research studies than is possible with human annotating alone. Currently, annotating problems by hand is burdensome. Our annotator required about 5 hours to apply our measurement scheme to $\sim 100$ student responses for one problem. In contrast, once trained the processing time for our models to annotate the same amount of data is less than 1 minute. Furthermore, training a model to perform this task is doable in less than 3 hours. We hypothesize that with much more labeled training data across many additional problems, the model proposed here could perform well on unseen problem instances in introductory mechanics. However, the amount of data needed to cross such a threshold is unclear. Regardless, initial efforts to improve generalization across problems while reducing required annotation time should be directed towards collaborative annotation as well as semi-supervised learning methods \cite{vanengelenSurveySemisupervisedLearning2020}. 

Training models to measure mechanistic sensemaking on more data from new problems opens the door for interesting studies that have been difficult for PER. Consider this example hypothesis: ``Evidence of mechanistic sensemaking early in a course is better than early canonical correctness  at predicting both sensemaking and correct understanding at the end of the course.'' Testing that hypothesis would involve many hours of human analysis; the ML we have begun to develop would be able to carry out this investigation quickly and efficiently. ML-enabled measurement could be used to conduct a randomized controlled experiment in PER, such as the one in Kuo et al.~\cite{PhysRevPhysEducRes.16.020109}, but specifically for understanding the variables that lead to more mechanistic sensemaking in students. It could also be used to evaluate various introductory physics problems as being of low or high quality depending on whether a large enough body of students displayed (or did not display) evidence of mechanistic sensemaking in their responses (e.g.~problem 1 in our dataset).

\subsection{Generalizability}
\label{sec:generalizability}

Russ et al.~\cite{mech_reason} developed a framework for assessing mechanistic reasoning that has been in use in PER for almost 20 years. To develop ML, we adapt the categories based on a particular dataset to produce a scheme with 7 domains. Researchers with similar datasets within the same conceptual domain of classical mechanics may find the scheme and the ML it supports useful as is. New contexts, however, may require new projections of Russ's et al's framework. For example, evidence of mechanistic reasoning in student explanations of electricity and magnetism or quantum mechanics will look different. It is likely that \emph{analogies} would be an essential domain for some subject matter. 

\subsection{Reliability}
\label{sec:reliability}
By designing and defining the 7 domains for our dataset, and providing annotated examples, we hoped to reduce ambiguity for human annotators - those who might look to reproduce results on our data or to apply the scheme to their own. To measure this, we conducted an inter-rater reliability (IRR) test. Two annotators who took smaller roles in development of the domains, one professor of computer science and one senior undergraduate student in computer science, used the scheme to measure the last 15 student responses across the four physics problems (60 student responses total). The first author provided them with a codebook containing: the domain definitions, the four problems, the criteria vectors for each problem, and 5 annotated student examples per problem with some explicit reasons as to why certain coding decisions were made. 
Annotators provided initial ratings, then engaged in a collaborative discussion to resolve ambiguity and clarify disagreement. After this discussion, new ratings were assessed.

Among the three annotators (2 new raters alongside the first author who conducted all manual annotation for this work), the average accuracy and Cohen's Kappa values \cite{Cohen1960} achieved pre-discussion are displayed in Table \ref{pre_IRR}. The values that emerged post a discussion are presented in Table \ref{post_IRR}.

\begin{table}[ht]
\centering
{ 
\begin{tabular}{lcc}
    \toprule
    & \textbf{Ave. Accuracy} & \textbf{Ave. Cohen's Kappa} \\
    \midrule
    \textbf{Problem 1} & 1.0 & 1.0 \\
    \textbf{Problem 2}  & 0.86 & 0.66 \\
    \textbf{Problem 3} & 0.82 & 0.63 \\
    \textbf{Problem 4} & 0.83 & 0.48 \\
    \bottomrule
\end{tabular}
\caption{Pre-Discussion}
\label{pre_IRR}
}

\begin{tabular}{lcc}
    \toprule
    & \textbf{Ave. Accuracy} & \textbf{Ave. Cohen's Kappa} \\
    \midrule
    \textbf{Problem 1} & 1.0 & 1.0 \\
    \textbf{Problem 2}  & 0.90 & 0.79 \\
    \textbf{Problem 3} & 0.94 & 0.88 \\
    \textbf{Problem 4} & 0.88  & 0.69 \\
    \bottomrule
\end{tabular}
\caption{Post-Discussion}
\label{post_IRR}

\end{table}

Problem 1 showed perfect agreement, but very few of the student examples show evidence of mechanistic sensemaking. (This suggests a form of utility for instruction: perhaps this is not a useful problem to assign.) For the remaining three problems, IRR scores were reasonably high ($>0.8$ accuracy). Cohen's Kappa values incorporate chance due to data imbalance and are lower across problems, showing moderate to substantial agreement according to the categories in \cite{Cohen1960}.

Discussion improved scores to  ${\geq}0.88$ accuracy, and all Cohen's Kappa values increased to indicate substantial and almost perfect agreement \cite{Cohen1960}. After discussion, our codebook language was updated to improve its clarity.

For all results in Sec.~\ref{deployability_analysis}, we trained ML to reproduce the labels provided by our single primary annotator.
The 60 annotations from 2 other raters studied here were not included in our ML model training, as the IRR procedure was conducted in response to helpful reviewer feedback on the draft post model training based on the first author's coding alone. As such, we highlight that this is a limitation of the current work - we do not know how our ML model will perform on measurements from multiple annotators nor do we have a model that corrects for mis-labeling. Investigating these directions is important for future larger-scale studies. Once an ML model is trained, the same student response will always receive the same annotation. Prediction is fully deterministic. 

\subsection{Sustainability $\&$ Privacy.}
As ML research advances, new encoders may be able to improve the overall classification performance. However, many of these models require High Performance Computing (HPC) access and large amounts of energy resources for training \cite{stochastic_parrots}. As language modeling in ML improves, it is worth benchmarking new methods while also taking into account the amount of compute resources required to obtain only modestly better performance. We only consider models which are open-access, which allows us to train classifiers for our specific task and does not require us to share student data with a 3rd party.

\section{Conclusion \& Future Work}
\label{sec:conclusion}

In this work, we contribute a PER-ML measurement scheme to quantify evidence of mechanistic sensemaking in student responses to introductory mechanics problems. By introducing an ML implementation, our scheme can be more efficiently applied to many student responses on many problems than manual coding. This can enable large-scale quantitative research studies in PER, though this technology has both promises and limitations (see Sec.~\ref{limitations}) and there is no ``magic bullet.''

In our work to date, we studied three model variants with different language encoders that require different levels of computational resources. We conducted a comparative analysis on real student responses. Results show that the ML models have good average performance on data from new students (0.916 AUROC for the best model). Performance does differ across domains of the scheme, in part reflecting the variance in the amount of available data per domain. 

Throughout the research process, we identified new open questions and challenges. We thus conclude with reflections that we hope will motivate further research. 

Recall we describe two steps of refinement from the prior framework of mechanistic reasoning ~\cite{mech_reason}. The first was to design a scheme for \emph{measuring} evidence of mechanistic reasoning in the context of student explanations in introductory mechanics. The second step was to train ML based on human annotated measurements for 100 students across four mechanics problems. Reflections below consider these steps in reverse order.

\emph{(1) How can the ML implementation of this measurement scheme be improved?} A natural first reaction to this question is \emph{more data}. 
We hypothesize that with more labeled training data across new problems, the supervised learning models proposed here will be able to achieve better performance on the current task \emph{and} generalize to unseen problems.  
Given collaborative annotation efforts, such a model could be realized more efficiently and would save all users time in the long run. This also means that our model performance will depend on strong inter-rater agreement and accounting for human error in labeling. Considering future methods to account for different types of label \emph{noise} is an interesting direction for future work. 

There are a few paths that we could take to improve performance. The first is to experiment with newer encoders from the rapidly-advancing state-of-the-art frontier \cite{zhao2024surveylargelanguagemodels}. 
 We could use the LLaMa encoder~\cite{touvron_llama_2023} previously used by Fussell et al.~\cite{PhysRevPhysEducRes.21.010128}. 
 For an even newer encoder, we could explore the open-access, open-source 7-billion parameter Mistral model~\cite{jiang2023mistral7b}. As off-the-shelf language models improve, one might not need as much task-specific training data to obtain quality performance, though it is hard to imagine doing well with less than our 4-problem, 100 student dataset. 
 Task-specific training or in-context learning \cite{brown2020languagemodelsfewshotlearners} will always be required. Out-of-the-box language encoders or encoder-decoder models (e.g. ChatGPT) are trained on general corpora of language, whereas the mechanistic sense-making of interest in PER reflects more particular disciplinary practices.   

Another approach to improve the ML would be to leverage the dependency structure across our domains. As a brief intuitive example, consider the perspective of an annotator. If the annotator identifies that the student response did not discuss even the central object in the objects domain, it is essentially almost certain that properties, spatial organizations, and movement activities are also not provided. This correlation between domains could be exploited for better modeling, and may provide valuable inductive bias for achieving quality performance with less labeled data. 

Lastly, we can pursue semi-supervised learning methods~\cite{vanengelenSurveySemisupervisedLearning2020}, which optimize model outputs to match labels that we do have, while also learning from patterns in a large, easier-to-collect un-labeled dataset  (in our case, student response text without criteria vector annotations). We believe each of these paths forward are potentially valuable directions for improvement. 

\emph{(2) How can new useful measurement schemes be achieved?} We set out, and we remain interested, to assess student reasoning for sensemaking, in particular sensemaking relevant to the disciplinary practices of physics. For the present effort, we limited our attention to mechanistic sensemaking, drawing on prior work in PER and working to connect it to progress in ML. We limited our attention further to introductory mechanics. Other areas such as electricity and magnetism or quantum mechanics, would need different measurement schemes for evidence of mechanistic sensemaking. We are not prepared to hypothesize that ML could create such schemes, but it would be worth exploration. There are also many other approaches to explore for incorporating insight from PER and ML research. There are other bases in the literature regarding evidence of sensemaking, broader than mechanistic. There are also other data modalities (e.g. hand-written solutions with diagrams, graphs, and sketches; audio and visual recordings). These choices will influence our ML decisions, and vice versa. Depending on the ML constraints (e.g. complexity and interpretability of the model, amount of data and compute resources needed), frameworks from the PER literature will need to be adapted. Thus, as shown in this work, we put forth that the research process that integrates PER and ML is \emph{interactive}. Ideas from the PER and ML side will influence one another until \emph{convergence} - which is why we want to encourage collaborations between longtime members of the PER community and ML researchers who are interested in bringing their domain expertise to advancing the mission of PER. It is really through these collaborations that we can develop meaningful schemes that take advantage of the efficiency promise of ML, and unlock new insights in understanding student thinking.


\begin{acknowledgments}
We are grateful for financial support from the U.S. National Science Foundation via awards GCR $\#2428640$ and CAREER $\#2338962$. We wish to acknowledge other members of the Hughes lab research group, namely Ethan Harvey and Cynthia Feeney, for thoughtful ML discussions.~Additionally, we'd like to thank the baristas at Jaho (Gabriel) and Starbucks (Ken) for supporting us through the final stages of this project. 
\end{acknowledgments}

\bibliography{ml_ed.bib,refs_from_zotero.bib}
\pagebreak
\appendix 
\section{Physics problems with criteria questions}\label{dataset_details}
We provide all four physics problems as they originally appeared to students from Ref.~\cite{textbook}. For each problem, we show the specific annotation scheme - i.e. the specific criteria questions for each problem. \\

\fbox{%
  \begin{minipage}{0.45\textwidth}
    \textbf{\underline{Problem 1}}
    \vspace{0.5em}
    
    \includegraphics[width=1\linewidth]{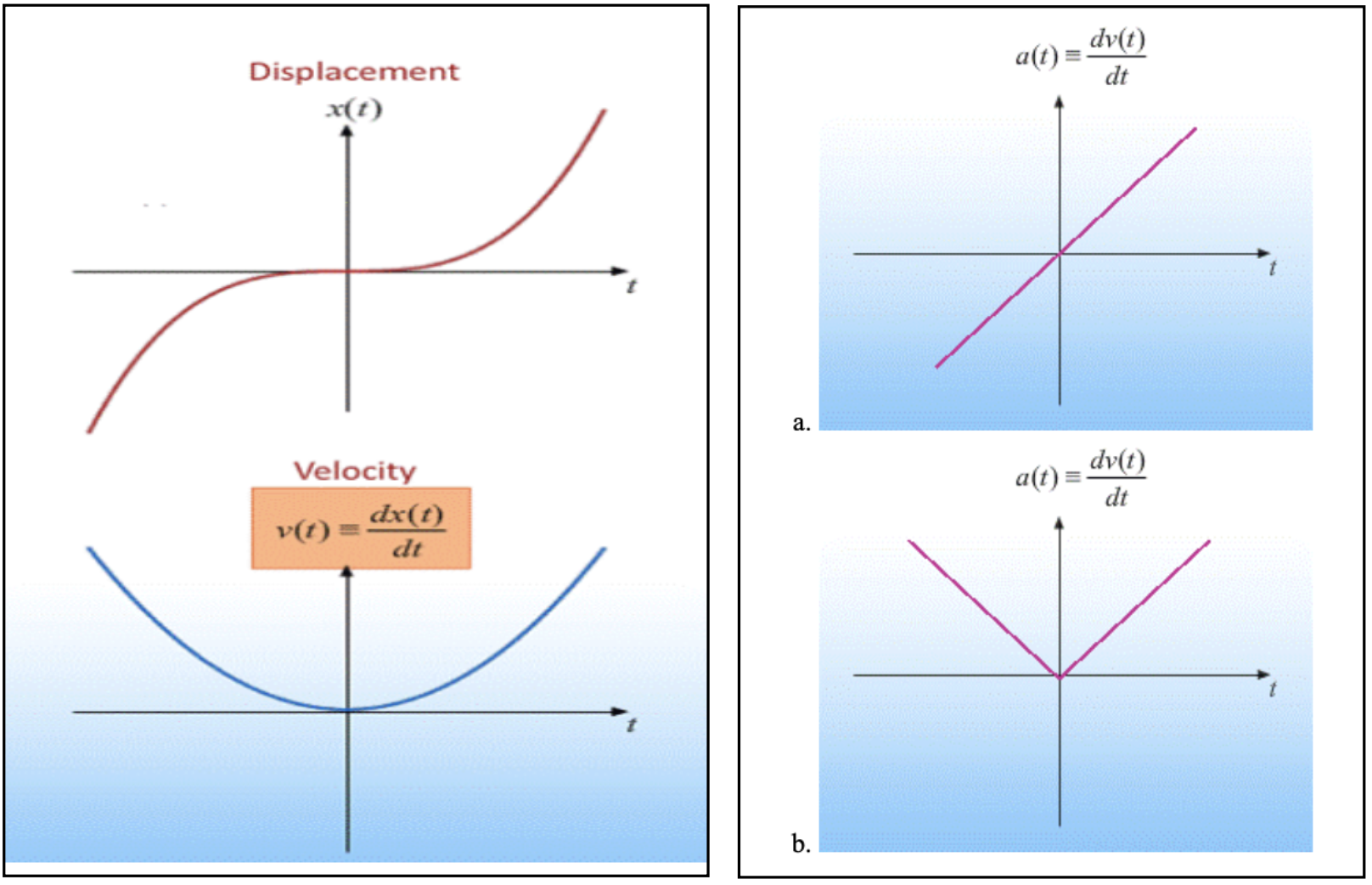}
    \vspace{0.5em}
    1. Which plot best represents the acceleration curve associated with the displacement and velocity curves shown here? \\

    2.~Briefly explain your answer to the previous question. \\

    \textbf{\underline{Annotation Procedure 1}} \\

    (1) Does the student identify the object (object, car, etc.)? \\
    (2) Does the student identify the object orientation (e.g. left or right)? \\
    (3) Does the student identify movement 1 of the object(e.g. slow down)? \\
    (4) Does the student identify movement 2 of the object (e.g. zero)? \\
    (5) Does the student identify movement 3 of the object (e.g. speed up)?  \\

  \end{minipage}%
}

\fbox{%
  \begin{minipage}{0.45\textwidth}
    \textbf{\underline{Problem 2}}
    \vspace{0.5em}
    
    \includegraphics[width=1\linewidth]{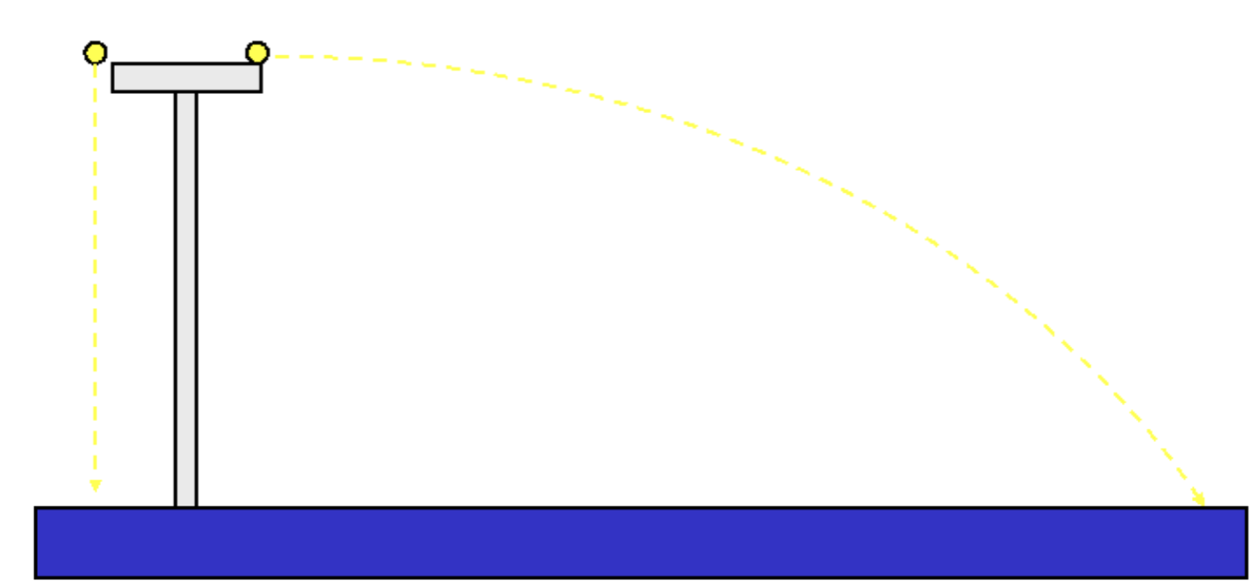}
    \vspace{0.5em}
    A physics demo launches a ball horizontally while dropping a second ball vertically at exactly the same time. \\ 
    1. Which ball hits the ground first? \\
    (a) Dropped ball \\
    (b) Launched (horizontally) ball \\
    (c) Both hit at the same time \\ 
    
    2. Briefly explain your answer to the previous question.

    \textbf{\underline{Annotation Procedure 2}} \\

    (1) Does the student identify object 1 (e.g. ball 1)? \\
    (2) Does the student identify object 2 (e.g. ball 2)? \\
    (3) Does the student identify the possible environmental influences on object 1 (e.g. gravity)? \\
    (4) Does the student identify the possible environmental influences on object 2 (e.g. gravity)? \\
    (5) Does the student identify object 1 spatial organization (e.g. down)? \\
    (6) Does the student identify object 2 spatial organization (e.g. horizontal up)? \\
    (7) Does the student identify object 1 motion post the drop (e.g. speeding up)? \\
    (8) Does the student identify object 2 motion post the drop (e.g. speeding up)? \\
    (9) Does the student identify the interaction between object 1 and gravity (e.g. gravitational force)? \\
    (10) Does the student identify the interaction between object 2 and gravity (e.g. gravitational force)? \\
    (11) Does the student identify the mechanistic relationship between gravity and a change in motion (e.g. Earth's gravitational force is a mechanism that causes the object to move down at a constantly increasing speed)? \\
    
  \end{minipage}%
}

\fbox{%
  \begin{minipage}{0.45\textwidth}
    \textbf{\underline{Problem 3}}
    \vspace{0.5em}
    
    \includegraphics[width=1\linewidth]{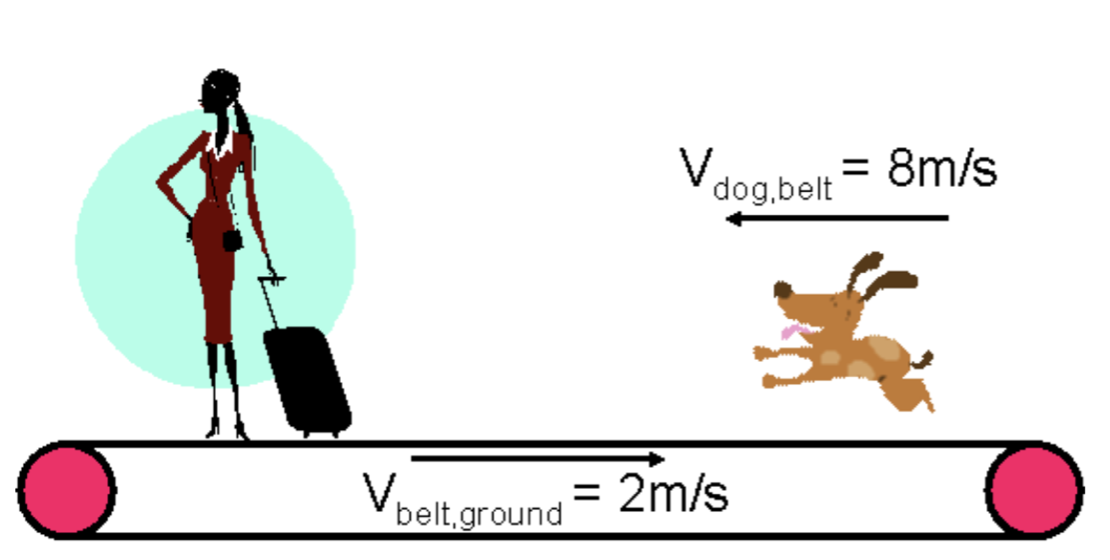}
    \vspace{0.5em}
    A girls stands on a moving sidewalk (conveyor belt) that is moving to the right at a speed of 2 m/s relative to the ground. A dog runs on the belt toward the girl at a speed of 8 m/s/ relative to the belt. 
     \\ 
    1. What is the speed of the dog relative to the girl? \\
    (a) 10 m/s \\
    (b) 6 m/s \\
    (c) 8 m/s \\ 

    \textbf{\underline{Annotation Procedure 3}} \\
    
    (1) Does the student identify object 1 (e.g. girl)? \\ 
    (2) Does the student identify object 2 (e.g. dog)? \\ 
    (3) Does the student identify object 3 (e.g. moving sidewalk, conveyor belt)? \\ 
    (4) Does the student identify the spatial organization of the dog (e.g. left)? \\
    (5) Does the student identify the spatial organization of the belt (e.g. right)? \\
    (6) Does the student identify the movement of the girl (e.g. at rest)?  \\
    (7) Does the student identify the movement of the dog (e.g. 8m/s constant speed)? \\ 
    (8) Does the student identify the movement of the belt (e.g. 2 m/s constant speed)? \\ 
    (9) Does the student identify the interaction between the girl and the belt (e.g. moving her to the right)?  \\
    (10) Does the student identify the interaction between the dog and the belt (e.g moving the dog to the right)? \\
    (11) Does the student identify the mechanistic relationship between motion of the observer and motion of the observed? (e.g. the observer moving relative to the object impacts the observed motion of the object) \\
        
  \end{minipage}%
}

\fbox{%
  \begin{minipage}{0.45\textwidth}
    \textbf{\underline{Problem 4}}

    \vspace{0.5em}
    1. The net force on a box is in the positive x direction. Which of the following statements best describes the motion of the box? \\
    (a) Its acceleration is parallel to the x axis. \\
    (b) Its velocity is parallel to the x axis. \\
    (c) Both its velocity and acceleration are parallel to the x axis. \\
    (d) Neither its velocity not its acceleration need to be parallel to the x axis. \\

    \textbf{\underline{Annotation Procedure 4}} \\

    (1) Does the student identify the object (e.g. object, box, etc.)? \\ 
    (2) Does the student identify the possible environmental influence acting on the object (e.g. applied net force)? \\ 
    (3) Does the student identify the object properties (e.g. mass)? \\ 
    (4) Does the student identify the object spatial organization before applied force (e.g. upwards)? \\
    (5) Does the student identify the object spatial organization after applied force (e.g. horizontal)? \\
    (6) Does the student identify the object movement prior to the applied force (e.g. constant velocity)? \\
    (7) Does the student identify the object movement after the applied force (e.g. final velocity)? \\
    (8) Does the student identify the interaction between the object and the net applied force (e.g. change in direction)? \\
    (9) Does the student identify the mechanistic relationship between an applied net force and a change in motion? (e.g. a net force applied in the x direction will be a mechanism that causes a change in motion/acceleration in the x direction)
        
  \end{minipage}%
}\\

\section{Annotation tables}\label{annotation_details} 

We provide an annotation summary. We showcase the number of student responses per problem, and the number of those explanations being labeled as positive for each specific criteria. 

\begin{table}[ht]
\centering
\begin{tabular}{cc}
\toprule
\textbf{Problem} & \textbf{Number of Responses} \\
\midrule
1 & 89 \\
2 & 98 \\
3 & 100 \\
4 & 98 \\
\bottomrule
\end{tabular}
\caption{The number of student responses per problem.}
\label{table:feature_vector_counts}
\end{table}

\begin{table}[ht]
\centering
\begin{tabular}{ccccc}
\toprule
\textbf{Criteria\#} & \textbf{Problem 1} & \textbf{Problem 2} & \textbf{Problem 3} & \textbf{Problem 4} \\
\midrule
1  & 9  & 89 & 97 & 66 \\
2  & 4  & 89 & 96 & 62 \\
3  & 6  & 49 & 78 & 8  \\
4  & 1  & 48 & 36 & 21 \\
5  & 4  & 40 & 9  & 46 \\
6  &    & 40 & 80 & 20 \\
7  &   & 46 & 92 & 44 \\
8  &    & 46 & 29 & 41 \\
9  &    & 21 & 54 & 42 \\
10 &    & 20 & 36 &    \\
11 &    & 18 & 13 &    \\
\bottomrule
\end{tabular}
\caption{The number of positive criteria labels per problem.}
\label{positive_label_summary}
\end{table}

\section{Loss function definition}\label{detailed_loss} 

Given the training dataset $D_{\text{Train}} =\{\textbf{x}^n, \textbf{y}^n\}_{n=1}^{N}$ of assumed independent student embeddings with human annotated labels, we train all parameters $\Theta$ to minimize the following loss function: 
\begin{align}
\mathcal{L}(\Theta) &= \mathcal{L}^{\text{BCE}}(\Theta) + \mathcal{R}(\Theta),
\\ \notag 
\mathcal{L}^{\text{BCE}}(\Theta) &=
\sum_{p=1}^{P} \sum_{i=1}^{N_{p}}\sum_{d=1}^{D=7}
\sum_{c \in C^p_{d}} \frac{\ell^{i,p}_{c,d}}{ D P N_p \max(1,|C^p_d|)},
\\ \notag 
\ell^{i,p}_{c,d} &=
 - y^{i,p}_{c,d}\log(\rho^{i,p}_{c,d}) - (1-y^{i,p}_{c,d})\log(1-\rho^{i,p}_{c,d}).
\end{align}

Here, $P$ is the total number of problems, $N_p$ is the total number of student responses per problem in the dataset $D_{\text{Train}}$, $D=7$ is the total number of domains, and $C^p_d$ is the set of all context criteria (indexed by $c$) in a particular domain $d$ for a specific problem $p$. In the binary cross-entropy expression, $y^{i,p}_{c,d}$ is the ground truth prediction for the $i^{th}$ student response associated with problem $p$ with respect to domain $d$ and specific criteria $c$. $\rho^{i,p}_{c,d}$ is the corresponding predictive probability. Note that by \emph{ground truth}, we mean the human-annotated labels provided in this work.  

The loss representing fit to the training data is averaged over all examples. The overall loss also includes a regularization term $\mathcal{R}(\Theta)$ over all weights. When we train \texttt{BERTfinetuned}, we have both a regularization term in the loss for classifier weights $\{\textbf{W}_d, b_{d}\}_{d=1}^D$ and the encoder weights of \emph{BERT} $\{\boldsymbol{\theta}\}$. We use an L2SP loss \cite{chelba2006adaptation, xuhong2018explicit}, which encourages weights for the LR classifiers to be close to zero, while encouraging BERT encoder parameters to be near pretrained values.

\section{Hyper-parameter strategy}\label{hyper-param-procedure} 

We outline our hyper-parameter procedure step by step such that the experiments in this work can be reproduced. For all experiments, we set the following seeds to control for randomness: $123$ when splitting the data, $360$ when running experiments, and $126$ when bootstrapping test sets. 

\textbf{\texttt{BoW} \& \texttt{BERTfrozen}.}  We randomly select $15$ students for our held-out test set ($60$ explanations) and use the rest of the data for a randomized 5-fold cross-validation procedure, where we stratify by student ID. The student ID stratification is to ensure that there is no data leakage between students at train and test time - i.e. we don't want the model learning student-specific correlations. As such, we ensure that all responses from a specific student are in the same fold. All splits are located in our Github Repo ``AuSeM''.

We perform a grid hyper-parameter search while optimizing for the AUROC score over the following: learning rates $\alpha = \{ 0.1, 0.01, 0.001, 0.0001, 0.00001, 0.000005, 0.000001\}$, and weight decays $\lambda_{1} = \{ 0.1, 0.01, 0.001, 0.0001, 1e-5, 0\}$. We average the value of AUROC from the last epoch from all 5 folds and pick the hyperparameter combination with the best AUROC score. The best model for \texttt{BoW} achieved an AUROC score of $0.875$ with the hyper-parameters: $\alpha = 0.01, \lambda_{1} = 0.001$. Similarly, for \texttt{BERTfrozen}, the best model achieved an AUROC score of $0.880$ with the hyper-parameters: $\alpha = 0.001, \lambda_{1} = 0.0001$. 

\textbf{\texttt{BERTfinetuned}.} We use the same held-out test set, but instead of conducting a cross-validation procedure which is computationally expensive for fine tuning, we use the rest of the data as a train/val set for the hyper-parameter selection. The train/val set are also stratified by student ID. The split is located in our Github Repo. ``AuSeM''.

We run a hyper-parameter grid search with an Adam optimizer for $1000$ epochs and batch-size to $N=32$ over the following: learning rate~$\alpha = \{0.1, 0.01, 0.001, 0.0001, 0.00001, 0.000001, 0.0000005,1e-7,1,1e-8\}$,~classifier weight decay~$\lambda_{1} = \{0,0.1, 0.01, 0.001, 0.0001, 0.00001\}$, and BERT encoder weight decay $\lambda_{2} = \{0,0.1, 0.01, 0.001, 0.0001, 0.00001\}$. We use two weight decay variables such that we can regularize the weights from \emph{BERT} separately from the weights of our LR, as we want to best preserve the information learned in the pre-trained model \cite{chelba2006adaptation, xuhong2018explicit}. This totals to 324 fine-tuning experiments for \texttt{BERTfinetuned} on Tufts' HPC services.

We sort all models by optimal final validation AUROC and select the model that achieves the highest final AUROC $= 0.946$. We observe that the model's behavior appears stable. This model's hyperparameters are: $\alpha = 1e-05$, $\lambda_1 = 0.0001$, and $\lambda_2 = 0.1$. Note that prior to sorting by final AUROC, we sorted by $\max$ AUROC to determine whether a reasonable early stopping point could be identified. While some models did have higher intermediate validation AUROC values, this was only the case for unstable models with too-high learning rates. As such, we determined them to be unreliable, and did not find a place where early stopping would be useful. 

\section{AUPRC Scores}\label{auprc_metrics} 

We show the test set AUPRC scores across domains for each classifier in Fig.~\ref{domain_bar_AUPRC}. The micro-averaged test set AUPRC score for each model is in Table \ref{table4}.

\begin{figure*}[!h]
  \centering
  \begin{subfigure}[b]{0.9\textwidth}\centering
\includegraphics[width=\linewidth]{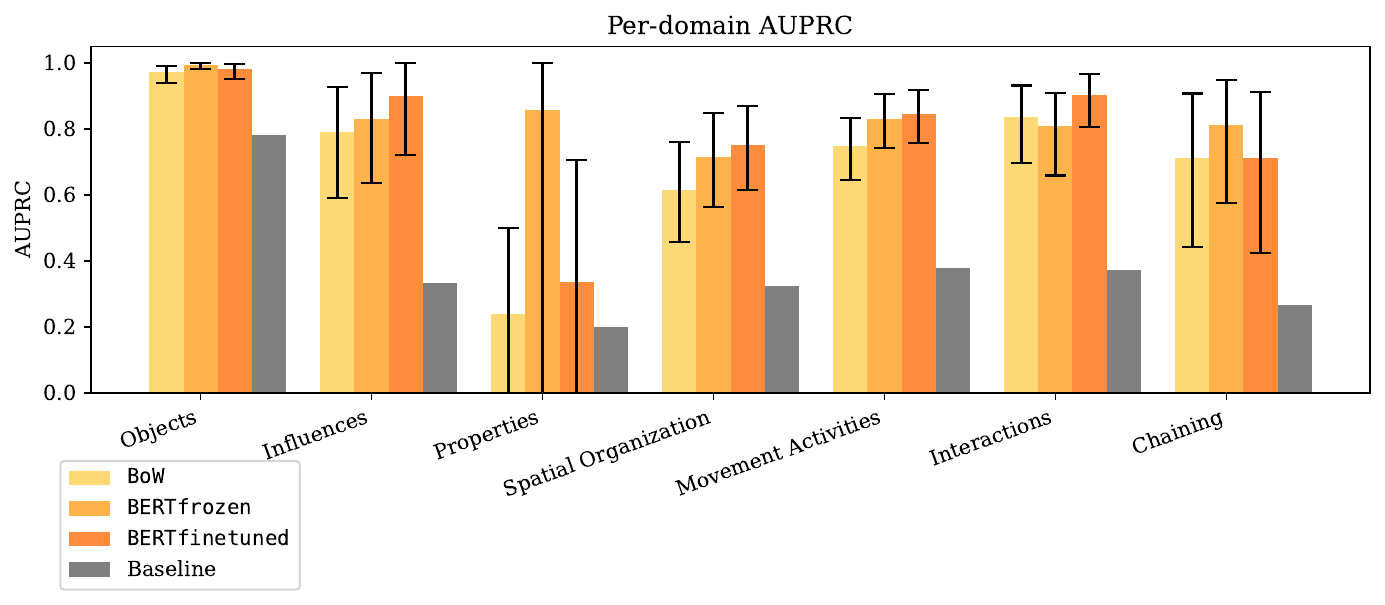}
  \end{subfigure}
  \caption{\textbf{AUPRC model performance across sensemaking domains.} For each domain, we show the average AUPRC and 95\% confidence interval from our 1000 bootstrapped test samples. \texttt{BERTfrozen} has the best performance in 3 domains (objects, properties, and chaining), and \texttt{BERTfinetuned} has a higher average AUPRC in the other 4 domains. All 3 models struggle in domains where there is limited data, especially in the properties domain where zero lies in the $95\%$ confidence interval. In grey we show the Baseline AUPRC value, the fraction of positive examples.}
\end{figure*}\label{domain_bar_AUPRC}

\begin{table}[ht]
\centering
{ 
\begin{tabular}{lcc}
    \toprule
    & \textbf{AUPRC (95\% CI)} & \textbf{Train time} \\
    \midrule
    \textbf{\texttt{BoW}} & 0.829 (0.786, 0.868) & 1 min \\
    \textbf{\texttt{BERTfrozen}}  & 0.876 (0.842, 0.906) & 7 min \\
    \textbf{\texttt{BERTfinetuned}} & 0.895 (0.862, 0.926) & 160 min \\
    \bottomrule
\end{tabular}
}
\caption{Avg.~model performance and comparison.}
\label{table4}
\end{table}

\section{Confusion matrices}\label{CONFUSION} 
Below, we present for each method a confusion matrix covering all yes/no criteria questions for the 15-student test set. To produce binary predictions, we threshold the predicted probability at a value of 0.5. We observe similar performance here as in the AUROC scores (which take all possible thresholds into account). \texttt{BERTfinetune} has the fewest classification errors overall. However, we see that \texttt{BERTfrozen} makes the same number of false positive mistakes (45). For deployment applications where one type of error has a larger cost than another, careful model and threshold selection are important.
\\

\noindent \textbf{\texttt{BoW}} on test set:
\[
\begin{array}{c|cc}
\text{Actual} \backslash \text{Predicted} & 0 & 1 \\
\hline
0 & 249 & 60 \\
1 & 66 & 165
\end{array}
\]

\noindent \textbf{\texttt{BERTfrozen}} on test set:
\[
\begin{array}{c|cc}
\text{Actual} \backslash \text{Predicted} & 0 & 1 \\
\hline
0 & 264 & 45 \\
1 & 64  & 167
\end{array}
\]

\noindent \textbf{\texttt{BERTfinetuned}} on test set:
\[
\begin{array}{c|cc}
\text{Actual} \backslash \text{Predicted} & 0 & 1 \\
\hline
0 & 264 & 45 \\
1 & 39  & 192
\end{array}
\]
\end{document}